\documentclass[letterpaper]{article} % DO NOT CHANGE THIS
\usepackage{aaai24}
\usepackage{times}  % DO NOT CHANGE THIS
\usepackage{helvet}  % DO NOT CHANGE THIS
\usepackage{courier}  % DO NOT CHANGE THIS
\usepackage[hyphens]{url}  % DO NOT CHANGE THIS
\usepackage{graphicx} % DO NOT CHANGE THIS
\usepackage{natbib}  % DO NOT CHANGE THIS AND DO NOT ADD ANY OPTIONS TO IT
\usepackage{caption} % DO NOT CHANGE THIS AND DO NOT ADD ANY OPTIONS TO IT
\usepackage{microtype}
\usepackage{graphicx}
\usepackage{caption}
\usepackage{enumitem}
\usepackage{amsmath}
\usepackage{amssymb}

\usepackage{latexsym}
\usepackage{multirow}
\usepackage{subfigure}

\usepackage{algorithm}
\usepackage{algorithmic}
\usepackage{newfloat}
\usepackage{listings}
\DeclareCaptionStyle{ruled}{labelfont=normalfont,labelsep=colon,strut=off} % DO NOT CHANGE THIS
\floatstyle{ruled}
\newfloat{listing}{tb}{lst}{}
\floatname{listing}{Listing}
\newcommand{\model}{TriSampler}
\newcommand{\smodel}{TriSampler\ }

\title{TriSampler: A Better Negative Sampling Principle for Dense Retrieval}
\author{
    Zhen Yang,
    Zhou Shao,
    Yuxiao Dong,
    Jie Tang\footnote{Jie Tang is the Corresponding Author.}
}
\affiliations{
    Department of Computer Science and Technology, Tsinghua University, Beijing, China \\
    yangz21@mails.tsinghua.edu.cn, \{shaozhou,yuxiaod,jietang\}@tsinghua.edu.cn 
}
\usepackage{bibentry}
\begin{document}

\maketitle

\begin{abstract}

Negative sampling stands as a pivotal technique in dense retrieval, essential for training effective retrieval models and significantly impacting retrieval performance. While existing negative sampling methods have made commendable progress by leveraging hard negatives, a comprehensive guiding principle for constructing negative candidates and designing negative sampling distributions is still lacking. To bridge this gap, we embark on a theoretical analysis of negative sampling in dense retrieval. This exploration culminates in the unveiling of \textit{the quasi-triangular principle}, a novel framework that elucidates the triangular-like interplay between query, positive document, and negative document. Fueled by this guiding principle, we introduce \textbf{\model}, a straightforward yet highly effective negative sampling method. The keypoint of \smodel lies in its ability to selectively sample more informative negatives within a prescribed constrained region. Experimental evaluation show that \smodel consistently attains superior retrieval performance across a diverse of representative retrieval models.

\end{abstract}

\section{Introduction}
Recently, dense retrieval has gained tremendous attention for its remarkable performance across a spectrum of real-world downstream applications, such as open-domain question answer~\cite{karpukhin2020dense}, web search~\cite{xiong2020approximate}, and conversational search~\cite{yu2021few}. 
Within the domain of dense retrieval, the focal point lies in the retrieval models' ability to effectively distinguish pertinent documents for a given query from the vast pool of non-relevant documents within the corpus.
This task, while crucial, is confronted with the challenge of managing an extensive set of negative documents. Attempting to leverage the entirety of these negatives is often impractical, which highlights the significance of negative sampling.

Indeed, negative sampling emerges as a vital technique in tackling this challenge. By advisably selecting a subset of negative documents, negative sampling allows for a more efficient and effective training process. This strategic management of negative samples not only mitigates the computational burden but also enhances the model's capacity to discern meaningful patterns, thereby contributing to the overall success of dense retrieval methods.

\begin{figure}[t]
    \centering    
    \vspace{-0.2cm} 
    \includegraphics[width=0.48\textwidth]{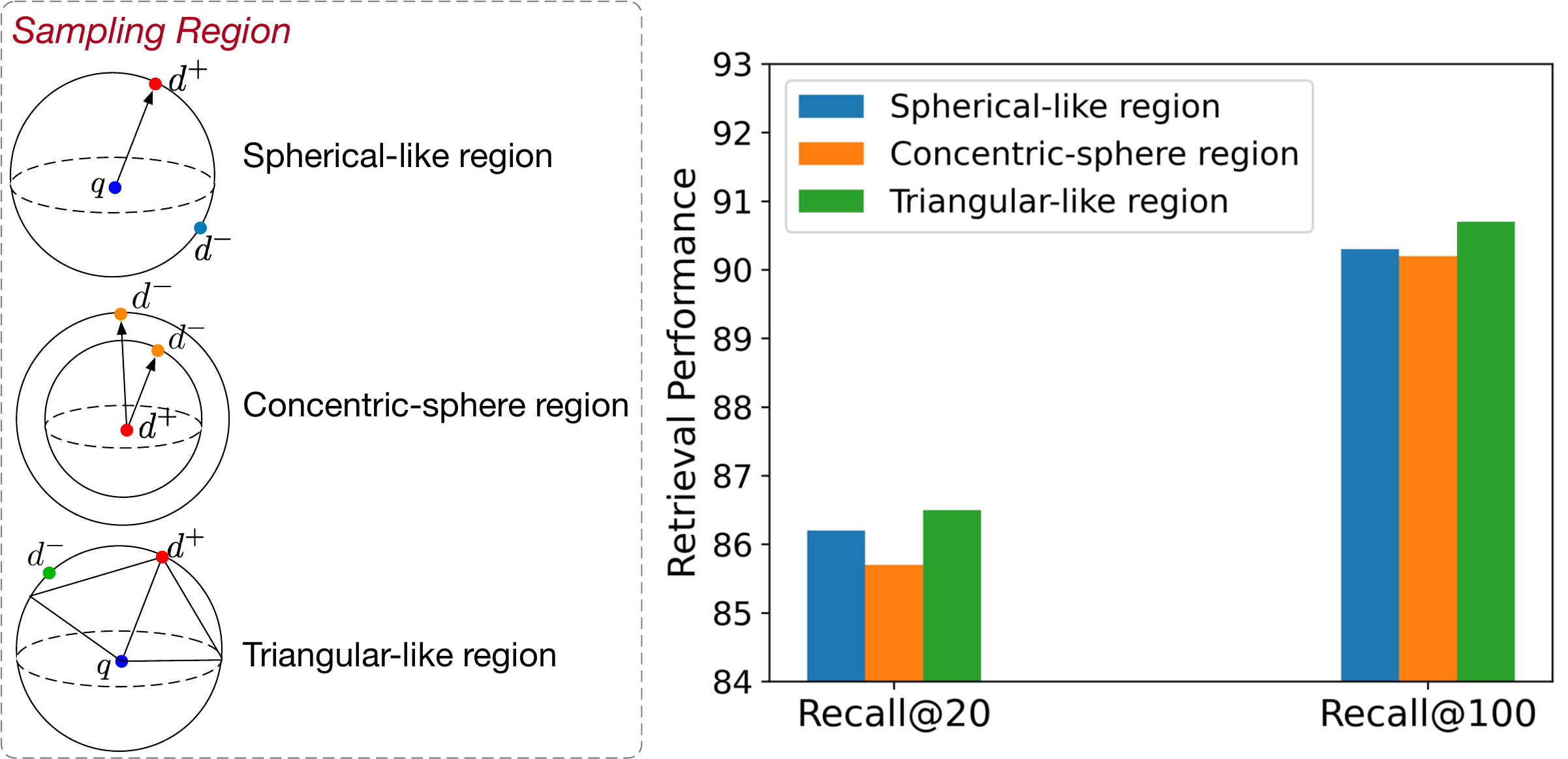}
    \caption{Insight experiments to illustrate the significance of the quasi-triangular principle.}
    \label{fig:insights}
    \vspace{-0.6cm}
\end{figure}

The pursuit of optimizing negative sampling techniques for dense retrieval has been a topic of considerable investigation. Previous efforts~\cite{karpukhin2020dense,kalantidis2020hard,zhan2021optimizing,qu2020rocketqa,chuang2020debiased} have delved into a range of methods to sample negatives at scale, encompassing strategies like in-batch negatives, random negatives, hard negatives, and debiased negatives. Drawing inspiration from the domain of contrastive learning~\cite{oord2018representation,he2020momentum,chen2020simple}, dense retrieval models adopt in-batch negatives. This approach, which represents a specialized instance of random negatives, streamlines training efficiency by reusing samples within the same batch, thus eliminating the need for additional sampling operations. Despite its merits, results in several works~\cite{faghri2017vse++,kalantidis2020hard,robinson2020contrastive,karpukhin2020dense,gao2021complement} indicate that such easy random negatives may not provide sufficient information for model training, potentially resulting in suboptimal retrieval performance. To address this issue, hard negative sampling methods~\cite{karpukhin2020dense,xiong2020approximate,zhan2021optimizing,qu2020rocketqa,sun2022reduce} have emerged as effective strategies for performance enhancement. These strategies focus on identifying and sampling top-k hard negatives based on the current model or leveraging an auxiliary retrieval model. However, the utilization of hard negatives presents a pivotal challenge: the risk of incorporating false negatives, which can detrimentally impact the retrieval performance~\cite{schroff2015facenet,chuang2020debiased,qu2020rocketqa,zhou2022simans}.

While prior studies have employed various negative sampling methods to achieve promising retrieval results, a general principle for guiding negative sampling is yet to be established. There is a distinct need for a clear and quantifiable approach that thoroughly elucidates that relationships among queries, positive documents, and negative documents. It is necessary to propose an explicit negative sampling methodology that adeptly selects more informative negative samples within a constraint region, effectively capturing the essential relationships.

Thus, we propose a general negative sampling principle called \textit{the quasi-triangular principle}, which mandates the constraint of sampled negatives within a triangular-like region. Specifically, this principle simulates the pairwise relationship among a training triple $(q, d^+, d^-)$, emphasizing the selection of negative samples from a region that mirrors the geometric characteristics of a triangle. To delve deeper into this negative sampling principle, we design two extended experiments: (1) sampling negatives from a spherical-like region centered on the query, with the radius defined by the similarity to the positive document; (2) sampling negatives within a concentric-sphere region centered on the positive document, aiming to find negatives that are related but not overly similar to it.
As depicted in Figure~\ref{fig:insights}, the outcome of constraining negatives within a triangular-like region notably elevates retrieval performance. This constrained sampling space empowers the retrieval model to more effectively distinguish relevant (positive) and irrelevant (negative) documents, bolstering the model's discriminative capabilities. Insights gleaned from these experiments suggest that confining negatives within a triangular-like region can augments retrieval performance.

To translate the innovative \textit{quasi-triangular principle} into actionable practice, we introduce a straightforward and efficient negative sampling methodology named \textbf{\model}. This method encapsulates a two-fold approach, involving the construction of negative candidates and the implementation of a specialized negative sampling distribution. The construction process is meticulously designed to capture the essence of the triangular-like relationships within the data, setting the foundation for an informative selection of negative samples. The implementation of distribution leverages two well-designed sampling distributions to guide the selection of negatives within the predefined triangular-like candidate region. The seamless integration of these two critical components solidifies \smodel as a general negative sampler.

\paragraph{Contributions:}  
In this paper, we present a general principle for negative sampling, referred to as the \textit{quasi-triangular principle}, centered on the idea of constraining sampled negatives within a triangular-like region that encapsulates the relationships between queries, positive documents, and negative documents. To implement this principle, we propose a simple and effective negative sampling method called \model. This methodology rests on two pillars: the construction of negative candidates and the implementation of a specialized negative sampling distribution. Empirical validation across four diverse retrieval benchmarks demonstrate that \smodel can achieve better retrieval performance compared to other negative sampling methods. 

\label{sec:intro}

\section{Related Work}

\paragraph{Dense retrieval.} 

Dense retrieval~\cite{lee2019latent,karpukhin2020dense, xiong2020approximate, khattab2020colbert} shows tremendous success in many downstream tasks (e.g. open-domain QA and web search) compared with the traditional sparse retrieval models (e.g. TF-IDF and BM25). The primary paradigm is to model semantic interaction between queries and passages based on the learned representations. 
Most dense retrieval models leverage the pretrained language models to learn latent semantic representations for both queries and passages. 
\citet{lee2019latent} first proposed the dual-encoder retrieval architecture based on BERT, paving the way for a new retrieval approach. In order to model fine-grained semantic interaction between queries and passages, Poly-encoder~\cite{humeau2019poly}, ColBERT~\cite{khattab2020colbert}, and ME-BERT~\cite{luan2021sparse} explored multi-representation dual-encoder to enhance retrieval performance. Besides, knowledge distillation has become a vital technique to enhance the capacity of the dual-encoder by distilling knowledge from a more powerful reader to a classical retriever~\cite{qu2020rocketqa,ren2021rocketqav2,lin2020distilling,hofstatter2021efficiently}.

Recently, massive works have investigated task-related pre-training methods for dense retrieval models~\cite{gao2021your,gao2021unsupervised,wang2021tsdae,ren2021pair,ouguz2021domain,meng2021coco}. Condenser~\cite{gao2021your} proposed the Condenser architecture to enforce the late backbone layers to aggregate the whole information. coCondenser~\cite{gao2021unsupervised} leveraged contrastive learning to incorporate a query-agnostic contrastive loss. PAIR~\cite{ren2021pair} and DPR-PAQ~\cite{ouguz2021domain} also designed special tasks in pre-training to enhance retrieval models. Additionally, jointly training retrieval models with the rerank model can bring about better performance. ~\citet{sachan2021end} proposed an end-to-end training method to jointly or individually model the retrieved documents. \citet{zhang2021adversarial} adopted adversarial training to model the retriever and the reranker.

\paragraph{Negative sampling in dense retrieval.} 

Several recent works~\cite{karpukhin2020dense,xiong2020approximate,qu2020rocketqa,zhan2021optimizing} demonstrate that hard negative sampling plays a crucial role in enhancing dense retrieval. Previous studies on negative sampling can be roughly categorized into three categories: (1) random sampling is the simplest way to obtain negatives. As an efficient random sampling method, in-batch negatives are widely used in dense retrieval models~\cite{karpukhin2020dense,zhan2021optimizing}. Such an approach is sub-optimal because random negatives have been proven to be too easy for learning effective models. RocketQA~\cite{qu2020rocketqa} adopted cross-batch negatives to increase the number of random negatives, resulting in better performance. (2) hard negative sampling can improve model generalization and accelerate convergence. DPR~\cite{karpukhin2020dense} additionally integrated hard negative passages from BM25 into in-batch negatives for dense passage retrieval. ANCE~\cite{xiong2020approximate} verified that global hard negatives obtained from the current retrieval model can significantly enhance the retrieval performance. ADORE~\cite{zhan2021optimizing} proposed a dynamic negative sampling method to train retrieval models. ANCE-Tele~\cite{sun2022reduce} combined past iterations by a momentum queue and future iterations by a lookhead operation to select hard negatives for stable training. (3) debiased hard negative sampling can efficiently alleviate false negatives. RocketQA~\cite{qu2020rocketqa} utilized a well-trained cross-encoder to select hard negatives for the dual-encoder training. SimANS~\cite{zhou2022simans} proposed ambiguous negatives to reweight the relevant score with the positives. Different from the abovementioned methods, our \smodel aims to sample negatives within a triangular-like region based on a general \textit{quasi-triangular principle}, which constraints the range of negative candidates and provides more informative negatives for model training.

\label{sec:realted}

\section{Understanding Negative Sampling}

In this section, we commence by providing an overview of the foundational concepts for dense retrieval. Subsequently, we delve into a comprehensive analysis of the pivotal role that negative sampling plays within the context of dense retrieval.

\subsection{Preliminary for Dense Retrieval}

Previous dense retrieval works~\cite{karpukhin2020dense,xiong2020approximate} aim to distinguish the most relevant documents $\mathcal{D}^+$ from a large document corpus $\mathcal{D}$ for a given query $q$. Typically, these retrieval models leverage negative sampling method to sample several negatives to substitute the entire corpus for model training, thus significantly reducing training costs. The objective function for dense retrieval can be simplified as:
\begin{equation}
    \mathcal{L} = \sum_{q} \sum_{d^+ \in \mathcal{D}^+} \sum_{d^- \in \mathcal{D}^-} l(s(\textbf{h}_q,\textbf{h}_{d^+}), s(\textbf{h}_q, \textbf{h}_{d^-})) 
    \label{eq:loss}
\end{equation}
where $l(\cdot)$ represents a loss function, such as cross entropy or hinge loss, $s(\cdot)$ denotes the dot product used to measure the similarity metric, $\textbf{h}_q$ and $\textbf{h}_d$ represent query embedding and document embedding that are encoded by a query encoder and a document encoder respectively. The pre-trained language models (PLMs) ~\cite{devlin2018bert,liu2019roberta,zhang2019ernie} serve as dual-encoder and the representations of the [CLS] token are leveraged as embeddings. In subsequent research endeavors, it has been observed that integrating the average of both the first and last layers embeddings confers a significant performance advantage~\cite{li2020sentence,su2021whitening}.

The construction of negative candidates $\mathcal{D}^-$ depends on either the current retrieval model or sparse retrieval model like BM25. Subsequently, the final negatives are sampled by applying distinct negative sampling distributions tailored to each specific approach.

\subsection{Analysis for Negative Sampling}

A representative dense retrieval model is trained on a set of training triples $\{(q, d^+, \{d^-\}_{i=1}^n)\}$ where $(q, d^+)$ is a positive query-document pair and $\{d^-\}_{i=1}^n$ are the sampled negative irrelevant documents. A conventional contrastive loss for dense retrieval can be formulated as:  
\begin{equation}
  \mathcal{L} = - \log \frac{exp(s^+)}{exp(s^+) + \sum_{i=1}^n exp(s_i^-)} 
\end{equation} 
where $s^+$ denotes the positive similarity score between the query and the corresponding positive document $s(\textbf{h}_q, \textbf{h}_{d^+})$, $s^-$ represents the negative similarity score between the query and a negative document $s(\textbf{h}_{q}, \textbf{h}_{d^-})$.

The gradient of the aforementioned contrastive loss can be separated into two distinct components concerning $s^+$ and $s_j^-$: 
\begin{equation}
\begin{aligned}
    \frac{\partial \mathcal{L}}{\partial s^+} &= -  \frac{\sum_{i=1}^n exp(s_i^-)}{exp(s^+)+\sum_{i=1}^n exp(s_i^-)}, \\
    \frac{\partial \mathcal{L}}{\partial s_j^-} &= \frac{exp(s_j^-)}{exp(s^+)+\sum_{i=1}^nexp(s_i^-)}
\end{aligned} 
\label{partial}
\end{equation}

According to Equation~\ref{partial}, it becomes evident that the gradient with respect to the negative document is proportional to the negative similarity score $exp(s_j^-)$. Notably, when employing randomly sampled negatives, their extremely low similarity scores result in gradients that are nearly negligible, rendering a minimal impact on model training. Conversely, negatives extracted from the top $K$ nearest irrelevant documents yield comparatively higher similarity scores, thereby expediting the convergence of the retrieval model. However, the gradient with respect to the positive document becomes bounded to a fixed value when the negative similarity scores significantly exceed the positive ones. This necessitates the imposition of specific constraints on the relationship between positives and negatives. An uncomplicated relationship, such as $s^+ \approx s^-$ when negatives are drawn from a spherical-like region, can be effective. Such strategically selected negatives infuse more informative signals into the training process, mitigating the occurrence of either zero or fixed-value gradients.

The above analysis clearly demonstrates that negatives adhering to the constraint region $s^+ \approx s^-$ contribute to the elimination of overly hard or overly easy negatives. Furthermore, our approach involves integrating the similarity score between the positive document and negative samples into the negative sampling process. The specific interplay among queries, positive documents, and negative documents will be expounded upon in the next Section.

\label{sec:theory}

\section{Method}

As elaborated in the abovementioned Section, a promising negative sampling method should adhere to the constraint $s^+ \approx s^-$, implying that negatives are optimally drawn from a spherical-like region. However, considering the expansive expanse of the entire spherical region, it becomes evident that negatives located far from the positive document might not yield valuable insights, given the model's inherent capability to distinguish between positive and negative documents. 

In response to this challenge, we introduce the \textit{quasi-triangular principle}, which strategically confines the sampled negatives within a delimited triangular-like region, effectively carving out a subregion of the larger spherical space. By adopting this principle, we narrow the scope of negative sampling, focusing on a region that balances the need for informative negatives with the practicality of model training. Based on this principle, we develop a simple and effective negative sampling method \model.

\subsection{The Principle of Negative Sampling}\label{sub:principle}

Here, we propose \textit{the quasi-triangular principle} to simulate the pairwise relationship inherent in a training triple $(q, d^+, d^-)$, ultimately enhancing negative sampling in dense retrieval. This principle operates by constraining the domain of sampled negatives within a triangular-like region, deviating from the broader spherical-like region that has traditionally been employed.

\begin{figure}[t]
    \centering
    \vspace{-0.2cm} 
    \includegraphics[width=0.47\textwidth]{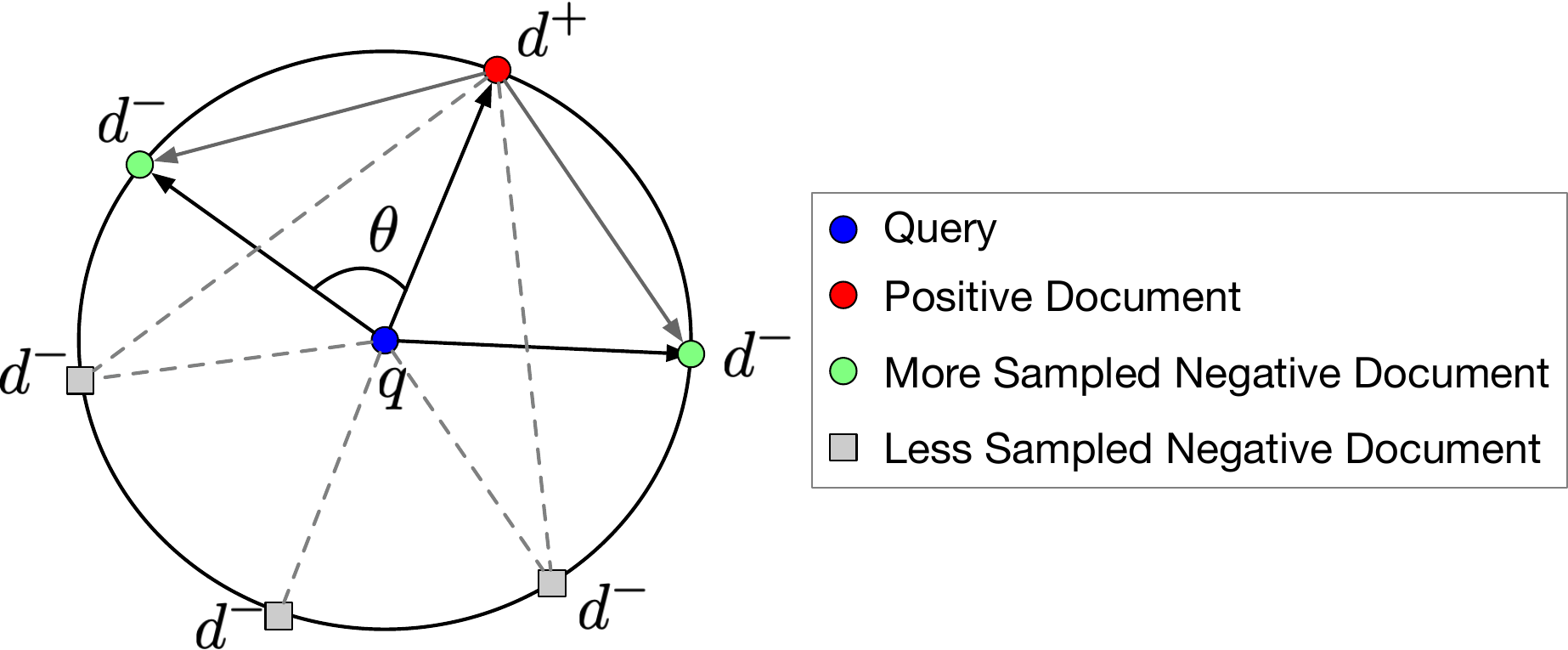}
    \caption{The proposed quasi-triangular principle for negative sampling in dense retrieval.}
    \label{fig:principle}
    \vspace{-0.5cm}
\end{figure}

The conceptual foundation of this principle is depicted in Figure~\ref{fig:principle}, which illustrates the planar projection of a sphere. In this context, the angular parameter $\theta$ characterizes the extent of the triangular-like region. Mathematically, $\theta$ is defined as:
\begin{equation}
\resizebox{.88\hsize}{!}{$
    \theta = |\arccos(\frac{s(\textbf{h}_{q}, \textbf{h}_{d^+})}{||\textbf{h}_{q}||\cdot ||\textbf{h}_{d^+}||}) -  \arccos(\frac{s(\textbf{h}_{q}, \textbf{h}_{d^-})}{||\textbf{h}_{q}||\cdot ||\textbf{h}_{d^-}||})$}
\end{equation}

Within the triangular-like region, the boundary for negatives is set at $\theta = 60^\circ$. This judicious constraint presents a noteworthy departure from the broader spherical-like region. This distinction serves to concentrate the sampled negatives in closer proximity to the positive documents. Consequently, the triangular-like region is thoughtfully designed to ensure that negatives exhibit a substantial degree of similarity with both the query and the positive document. This strategic arrangement proves pivotal in tackling issues associated with two key types of negatives: those that are too close to the query (potentially leading to false negatives) and those situated far from both the query and the positive (often resulting in uninformative negatives).

\subsection{Negative Candidates}
In order to enhance the informativeness of negative candidates, we seamlessly integrate the \textit{quasi-triangular principle} into the process of constructing negative candidates, denoted as $\mathcal{D}_q^-$. This is carried out within the bounds of the established triangular-like region, specific to each query $q$. Specifically, we first sample the top-ranked irrelevant documents in terms of the query based on the current retrieval model, which is widely used in previous hard negative selection methods~\cite{xiong2020approximate,zhan2021optimizing,zhang2021adversarial,zhou2022simans}. Subsequently, we determine the relevant scores between the positive document and the selected top-ranked irrelevant documents. Building upon the aforementioned relevant scores $s(\textbf{h}_{q}, \textbf{h}_{d^-})$ and $s(\textbf{h}_{d^+}, \textbf{h}_{d^-})$ where $d^- \in \textbf{TopK}_{s(q,\mathcal{D}^-)}$, we establish criteria for the construction of a more informative negative candidate set in alignment with the \textit{quasi-triangular principle}. This involves two pivotal constraints: 
\begin{itemize}[leftmargin=*]
    \item Negative candidates should conform to the first range constraint $s(\textbf{h}_{q}, \textbf{h}_{d^+}) \approx s(\textbf{h}_{q}, \textbf{h}_{d^-})$, effectively filtering out  too hard or too easy negatives; 

    \item  Negative candidates should be in line with the second range constraint $s(\textbf{h}_{d^+}, \textbf{h}_{d^-}) \geq s(\textbf{h}_{q}, \textbf{h}_{d^-})$, fostering the inclusion of negatives that are notably informative.      

\end{itemize}

\subsection{Negative Sampling Distribution}

The primary goal of negative sampling method is to design an effective distribution that effectively samples high-quality negatives from the negative candidates. Based on \textit{the quasi-triangular principle}, we formulate the first distribution to adhere to the range constraint $s(\textbf{h}_{q}, \textbf{h}_{d^+}) \approx s(\textbf{h}_{q}, \textbf{h}_{d^-})$. This distribution takes the form:
\begin{equation}
p_{d^-}^{(q)} \propto exp( -\frac{1}{4} * (s^- - s^+)^2)
 \label{eq: distribution}
\end{equation}
where $s^-$ and $s^+$ represent $s(\textbf{h}_{q}, \textbf{h}_{d^-})$ and $s(\textbf{h}_{q}, \textbf{h}_{d^+})$ respectively. This distribution embodies the aim of mitigating excessively hard negatives (i.e., false negatives) while concurrently reinforcing the first range constraint $s^+ \approx s^-$. Implementing this distribution exclusively over the top-ranked negative candidates $\textbf{TopK}_{s(q,\mathcal{D}^-)}$ yields the transitional negatives $\mathcal{\tilde{D}}_q^-$.

Turning to the second range constraint, we devise a novel distribution to sample the final negatives for the training of the retrieval model. Specifically, we allocate higher sampling probabilities to negatives that are in close proximity to the positive document within the triangular-like region. This distribution can be succinctly represented as:
\begin{equation}
    p_{d^-} \propto \text{ReLU}(s(\textbf{h}_{d^+}, \textbf{h}_{d^-})- s(\textbf{h}_{q}, \textbf{h}_{d^-}))
\end{equation}
where this distribution is conducted on transitional negatives $\mathcal{\tilde{D}}_q^-$.

The key insight of using the RuLU function is that it can exclude negatives that are not in the triangular-like region and further guarantee that negatives that are closer to positive possess higher sampling probabilities. By applying these distributions, we holistically adhere to the \textit{quasi-triangular principle}, successfully selecting a set of informative negative samples for retrieval model training.

\subsection{Discussion}

Algorithm~\ref{pipeline} represents the overall training process of \model. In the ensuing discussion, we delve into the connection and discrimination between \smodel and previous negative sampling methods.

\begin{algorithm}[t]
    \caption{Algorithm of \model}
    \label{pipeline}
    \textbf{Input}: Positive query-documents $\{(q, \mathcal{D}^+)\}$, document corpus $\mathcal{D}$. \\ 
    \textbf{Output}:  Negative documents $\mathcal{\hat{D}}_q^-$. 
    \begin{algorithmic}[1] %[1] enables line numbers
    \STATE Build ANN index on $\mathcal{D}$. \\
    \STATE Generate the top-ranked negative candidates $\textbf{TopK}_{s(q,\mathcal{D}^-)}$ from $\mathcal{D}$. \\
    \STATE  Sample transitional negatives $\mathcal{\tilde{D}}_q^-$ from $\textbf{TopK}_{s(q,\mathcal{D}^-)}$ with  distribution $p_{d^-}^{(q)}$. \\
    \STATE Sample final negatives $\mathcal{\hat{D}}_q^-=\{d^-\}_{i=1}^n$ based on distribution $p_{d^-}$ from $\mathcal{\tilde{D}}_q^-$. \\ 
    \end{algorithmic}
\end{algorithm}

\begin{itemize}[leftmargin=*]
    \item \textbf{\smodel vs RandNS.} RandNS~\cite{huang2020embedding} is a basic method that randomly samples negatives from a huge set of negative candidates. \smodel relies on the \textit{quasi-triangular principle} to sample more informative negatives within the triangular-like region. Different from RandNS that assigns equal weights for each negative, \smodel leverages a well-designed distribution to sample negatives. 

    \item \textbf{\smodel vs TopNS.} TopNS aims to sample top-k ones from all ranked negatives based on a dynamic-trained dense retrieval model~\cite{xiong2020approximate,zhan2021optimizing} or a sparse retrieval model~\cite{karpukhin2020dense} (BM25). Unlike TopNS which has a higher risk of false negatives, \smodel eliminates too hard negatives via a constraint triangular-like region.

    \item \textbf{\smodel vs SimANS.} SimANS~\cite{zhou2022simans} designs a negative sampling distribution to sample ambiguous negatives, which avoids sampling negatives that are either too hard or too easy. Similar to SimANS, \smodel also devises two distributions for the constraint region. The main difference between these is that \smodel limits negatives within a triangular-like region while SimANS leverages top-ranked negatives as the sampling region.    

    \item \textbf{\smodel vs ANCE-Tele.} ANCE-Tele~\cite{sun2022reduce} combines three types of negatives (standard ANCE negatives, momentum negatives, and lookahead negatives) to form negative candidates and then randomly sample negatives from the above candidates. Different from ANCE-Tele, \smodel constraints the sampling region within a triangular-like region and employs two specifically-designed distributions for sampling.
    
\end{itemize}

\label{sec:method}

\section{Experiments} 

\subsection{Experimental Setup}
\paragraph{Datasets.}
We conduct experiments on the first retrieval stage of four benchmarks: three passage retrieval datasets: MS MARCO passage (MS Pas)~\cite{nguyen2016ms},  Natural Questions (NQ)~\cite{kwiatkowski2019natural}, and TriviaQA (TQA)~\cite{joshi2017triviaqa}, and a document retrieval dataset: MS MARCO document (MS Doc)~\cite{nguyen2016ms}. The statistics of each dataset is illustrated in Table~\ref{tab:datasets}.

\begin{table}[h]
    \centering
    \small
    \renewcommand{\arraystretch}{1.}
    \setlength{\tabcolsep}{2mm}{
    \begin{tabular}{c|cccc}
        \hline
        Datasets & Training & Dev & Test & Documents \\
        \hline 
        NQ & 58,880 & 8,757 & 3,610 & 21,015,324 \\
        TQA & 60,413 & 8,837 & 11,313 & 21,015,324 \\
        MS Pas & 502,939 & 6,980 & - & 8,841,823 \\
        MS Doc & 367,013 & 5,193 & - & 3,213,835 \\
        \hline
    \end{tabular}}
    \caption{The statistics of four retrieval datasets.}
    \label{tab:datasets}
    \vspace{-0.4cm}
\end{table}

\paragraph{Evaluation metrics.} We evaluate retrieval performance using official evaluation methodologies, such as MMR$@10$ and R$@k$. For the NQ and TQA datasets, R$@20$ and R$@100$ serve as metrics to measure whether the top-$k$ retrieved passages contain the answer span. We evaluate the results on their dev datasets in terms of MRR$@10$ and R$@50$ for MS Pas dataset, MRR$@10$ and R$@100$ for MS Doc dataset.

\paragraph{Baselines.} We compare \model with previously established baselines for retrieval benchmarks. Baselines can be generally divided into the following categories. 

\begin{itemize}[leftmargin=*]
    \item \textbf{Sparse Retrieval.} The compared sparse retrieval models contains BM25~\cite{yang2017anserini} and  improved variants of BM25 models that incorporate pretrained language models, such as doc2query~\cite{nogueira2019doc2query}, DeepCT~\cite{dai2019deeper}, docTTTTTquery~\cite{nogueira2019document}, and GAR~\cite{mao2020generation}.   

    \item \textbf{Dense Retrieval.} Massive dense retrieval baselines have investigated a variety of training methods to improve the retrieval performance, such as hard negative sampling~\cite{karpukhin2020dense,xiong2020approximate,zhan2021optimizing,zhou2022simans}, distillation~\cite{qu2020rocketqa,lu2022ernie,ren2021rocketqav2}, integrating rerankers into retrievers~\cite{zhang2021adversarial}, pre-training~\cite{ren2021pair, gao2021unsupervised,gao2021your}, etc. Among these, hard negative sampling is a particularly important strategy. DPR~\cite{karpukhin2020dense}, RocketQA~\cite{qu2020rocketqa}, ANCE~\cite{xiong2020approximate}, ADORE~\cite{zhan2021optimizing}, and SimANS~\cite{zhou2022simans} attempt to design various negative sampling methods to obtain top-k hard negatives. 
    
\end{itemize}

\paragraph{Implementation details.} We implement \smodel based on SOTA dense retrieval model AR2~\cite{zhang2021adversarial} and run all experiments on 8 NVIDIA Tesla A100 GPUs. Following AR2, ERNIE-2.0-base~\cite{sun2020ernie} serves as a backbone model to encode queries and passages. Similar to SimANS~\cite{zhou2022simans}, we directly utilize checkpoints in the AR2 model to continue training with the proposed \model. For MS Doc dataset, the model parameters are initialized with STAR~\cite{zhan2021optimizing}. In our experiments, the ratio of positive to negative pairs is set to $1:15$, the inner product is leveraged to estimate the relevance score and Faiss~\cite{johnson2019billion} is adopted for efficient similarity search. We utilize the top-200 passages for NQ and TQA datasets and the top-400 documents for MS Pas and MS Doc datasets as negative candidates.

\begin{table*}[t!]
    \centering
    \renewcommand{\arraystretch}{1.}
    \setlength{\tabcolsep}{2mm}{
    \begin{tabular}{c|cc|cc|cc}
        \hline
      \multirow{2}{*}{Method} & \multicolumn{2}{c|}{NQ} & \multicolumn{2}{c|}{TQA} & \multicolumn{2}{c}{MS Pas} \\
         & R@20 & R@100 & R@20 & R@100 & MRR@10 & R@50\\
        \hline 
        BM25~\cite{yang2017anserini} & 59.1 & 73.7 & 66.9 & 76.7 & 18.7 & 59.2 \\
        doc2query~\cite{nogueira2019document} & - & - & -& - & 21.5 & 64.4  \\
        DeepCT~\cite{dai2019deeper} & - & -  & - & - & 24.3 & 69.0  \\
        docTTTTTquery~\cite{nogueira2019doc2query} & & & & & 27.7 & 75.6  \\
        GAR~\cite{mao2020generation} & 74.4 & 85.3 & 80.4 & 85.7 & - & - \\
        \hline
        DPR~\cite{karpukhin2020dense} & 78.4 & 85.4 & 79.3 & 84.9 & - & - \\
        ME-BERT~\cite{luan2021sparse} & - & - & - & - & 33.8 & -  \\
        Joint top-k~\cite{sachan2021end} & 81.8 & 87.8 & 81.3 & 86.3 & - & - \\
        Individual top-k ~\cite{sachan2021end} & 84.0 & 89.2 & 83.1 & 87.0 & - & -  \\        RocketQAv2~\cite{ren2021rocketqav2} & 83.7 & 89.0 & - & - & 38.8 & 86.2  \\
        PAIR~\cite{ren2021pair} & 83.5 & 89.1 & - & -  & 37.9 & 86.4 \\
        DPR-PAQ~\cite{ouguz2021domain} & 84.0 & 89.2 & - & - & 31.1 & - \\
        Condenser~\cite{gao2021your} & 83.2 & 88.4 & 81.9 & 86.2 & 36.6 & - \\
        coCondenser~\cite{gao2021unsupervised} & 84.3 & 89.0 & 83.2 & 87.3  & 38.2 & - \\
        ANCE-Tele~\cite{sun2022reduce} & 84.9 & 89.7 & 83.4 & 87.3 & 39.1 & - \\
        ERNIE-Search~\cite{lu2022ernie} & 85.3 & 89.7 & - & - & 40.1 & - \\
        AR2+SimANS~\cite{zhou2022simans} & 86.2 & 90.3 & 84.6 & 88.1 & 40.9 & 88.7 \\
        ColBERTv2~\cite{santhanam2021colbertv2} & - & - & - & - & 39.7 & 86.8 \\
        SimLM~\cite{wang2022simlm}  & 85.2 & 89.7 &  - & - & 41.1  & 87.8  \\
        LexMAE-Stage1~\cite{shen2022lexmae}  & - & - & - & -& 39.3 & -    \\
        LexMAE-Stage1~\cite{shen2022lexmae} & - & - & - & - & 40.8  & -  \\
        LexMAE~\cite{shen2022lexmae}  & - & - & - & - & 42.6 & -  \\
        \hline
        \hline
        ANCE~\cite{xiong2020approximate} & 81.9 & 87.5 & 80.3 & 85.3 & 33.0 & 81.1  \\
        ANCE + \smodel &\textbf{83.8} & \textbf{89.1} &\textbf{83.4}  &\textbf{87.2}  & \textbf{35.8} & \textbf{83.4}  \\
        \hline
        \hline
        RocketQA~\cite{qu2020rocketqa} & 82.7 & 88.5 & - & - & 37.0 & 85.5 \\
        RocketQA + \smodel  & \textbf{85.3} & \textbf{89.6}  & -  & - & \textbf{38.3} & \textbf{86.0}  \\
        \hline
        \hline
        AR2~\cite{zhang2021adversarial} & 86.0 & 90.1 & 84.4 & 87.9 & 39.5 & 87.8  \\
        AR2 + \model & \textbf{86.5} & \textbf{90.7} & \textbf{85.0} & \textbf{88.5} & \textbf{41.4} & \textbf{89.1} \\ 
        \hline
    \end{tabular}}
    \caption{Results on three retrieval benchmarks, including NQ test set, TQA test set, and MS Pas dev set. The results of baselines are directly obtained from the original papers and results not provided are marked as ``-''.}
    \label{tab:main_results}
    \vspace{-0.5cm}
\end{table*}

\subsection{Overall Results}

\smodel achieves a better retrieval performance than most of baselines on all metrics (See Table~\ref{tab:main_results} and Table~\ref{tab:result_ms_doc}). The improvements primarily stem from the superiority of the \textit{quasi-triangular principle} over previous hard negative sampling methods. Since the measurement principle between query-negatives and pos\_passage-negatives may share a \textit{quasi-triangular principle}, previous methods are unable to capture this principle or even overlook the impact of pos\_passage-negatives. 
Compared with pre-training methods for dense retrieval (SimLM and LexMAE), our TriSampler outperforms SimLM but not surpass LexMAE. LexMAE encompasses three stages: BM25 Negatives, Hard Negatives, Reranker-Distilled. While our TriSampler combined with AR2 exceeds the performance of LexMAE in its first two stages, it falls short in matching the effectiveness of LexMAE's complete three-stage protocol. This difference is primarily attributed to LexMAE's utilization of an advanced off-the-shelf reranker, a component that significantly enhances its overall performance. 
Experimental results in Table~\ref{tab:main_results} show that \smodel is a general method that can be naturally applied to various dense retrieval models. Such a method can provide more informative negatives to consistently improve downstream performance in dense retrieval.

\begin{table}[h]
    \centering
    \small
    \renewcommand{\arraystretch}{1.}
    \setlength{\tabcolsep}{2mm}{
    \begin{tabular}{c|cc}
        \hline
        Method & MRR@100 & R@100 \\
        \hline 
        BM25~\cite{yang2017anserini} & 27.9 & 80.7 \\
        DPR~\cite{karpukhin2020dense} & 32.0 & 86.4 \\
        ANCE~\cite{xiong2020approximate}  & 37.7 & 89.4 \\
        STAR~\cite{zhan2021optimizing} & 39.0 & 91.3 \\
        ADORE~\cite{zhan2021optimizing} & 40.5 & 91.9 \\
        AR2~\cite{zhang2021adversarial} & 41.8  & 91.4 \\
        AR2+SimANS~\cite{zhou2022simans} & 43.1 & 92.3 \\
        \hline 
        AR2+\model & \textbf{43.8} & \textbf{93.1} \\
        \hline
    \end{tabular}}
    \caption{Experimental performance on MS Doc dev set.}
    \label{tab:result_ms_doc}
    \vspace{-0.6cm}
\end{table}

\subsection{Why \smodel Performs Better?}

\paragraph{Perspective of candidates.}

To deepen the understanding of \model, we vary the selection methods of negative candidates and conduct two extended experiments on the NQ dataset and the MS Pas dataset using the AR2 retrieval model: 
(1) top-k query-document ranked negative candidates $\mathcal{D}_q^- = \textbf{TopK}_{s(q,\mathcal{D}^-)}$; (2) top-k document-document ranked negative candidates $\mathcal{D}_q^- = \textbf{TopK}_{s(d^{+},\mathcal{D}^-)}$.

As shown in Table~\ref{tab: candidates}, \smodel surpasses all other variants of negative candidate selection methods, indicating the effectiveness of \model. For $\textbf{TopK}_{s(q,\mathcal{D}^-)}$ and $\textbf{TopK}_{s(d^+,\mathcal{D}^-)}$, they seem to only account for the impact of the query or positive document on negatives, ignoring the triangular-like relationship outlined in Section~\ref{sub:principle}. \smodel combines these two methods based on \textit{the quasi-triangular principle}, which alleviates the excessive reliance on the query and constrains the region of negative candidates. Consequently, \smodel can achieve enhanced performance, suggesting that the triangular-like relationship is a valuable constraint for selecting negative candidates.

\begin{table}[h]
    \centering
    \small
    \renewcommand{\arraystretch}{1.}
    \setlength{\tabcolsep}{2mm}{
    \begin{tabular}{c|cc|cc}
        \hline
        \multirow{2}{*}{Method} & \multicolumn{2}{c|}{NQ}  & \multicolumn{2}{c}{MS Pas} \\
         & R@20 & R@100  & MRR@10 & R@50\\
        \hline 
        $\textbf{TopK}_{s(q,\mathcal{D}^-)}$ & 86.2 & 90.3 & 40.9 & 88.7 \\
        $\textbf{TopK}_{s(d^{+},\mathcal{D}^-)}$ & 85.5 & 90.4 & 40.3 & 88.5  \\
        \hline
        \model & 86.5 & 90.7 & 41.4 & 89.1 \\
        \hline
    \end{tabular}}
    \caption{Various negative candidate selection methods on the NQ dataset and the MS Pas dataset.}
    \label{tab: candidates}
    \vspace{-0.5cm}
\end{table}

\paragraph{Perspective of distributions.} 
To demonstrate the effectiveness of the negative sampling distribution proposed in \model, we evaluate the retrieval performances on three variations of \smodel 
on the MS Pas dataset: (1) Uniform sampling that assigns negative candidates with equal weights; (2) TopK Sampling that leverages the relevant score as sampling weights; (3) Debiased Sampling that computes sampling weights by reducing the impact of the positive relevant score. Table~\ref{tab: distribution} reveals that \smodel outperforms the other variant negative sampling distributions. According to Equation~(\ref{eq: distribution}), the negative sampling distribution suggested by \smodel adheres to \textit{the quasi-triangular principle}. This principle allocates higher sampling probabilities to negatives that are closer to the positive document within a restricted region. This observation confirms that a well-designed sampling distribution can indeed contribute to enhanced performance.

\begin{table}[h]
    \centering
    \renewcommand{\arraystretch}{1.15}
    \setlength{\tabcolsep}{1.5mm}{
    \begin{tabular}{c|cccc}
        \hline
         Method & MRR@10  & R@50 & R@1k \\
        \hline 
        Uniform Sampling & 39.7 & 87.9 & 98.6 \\
        TopK Sampling & 40.6  & 88.6  & 98.7  \\
        Debiased Sampling & 41.1 & 88.9 & 98.8 \\
        \hline
        \model & 41.4 & 89.1 & 98.9 \\
        \hline
    \end{tabular}}
    \caption{Various negative sampling distributions on the MS Pas dataset.}
    \label{tab: distribution}
    \vspace{-0.5cm}
\end{table}

\subsection{Further Analysis}

\begin{figure}[h!]
\centering
\vspace{-0.2cm} 
    \subfigure[NQ]{
    \includegraphics[width=0.22\textwidth]{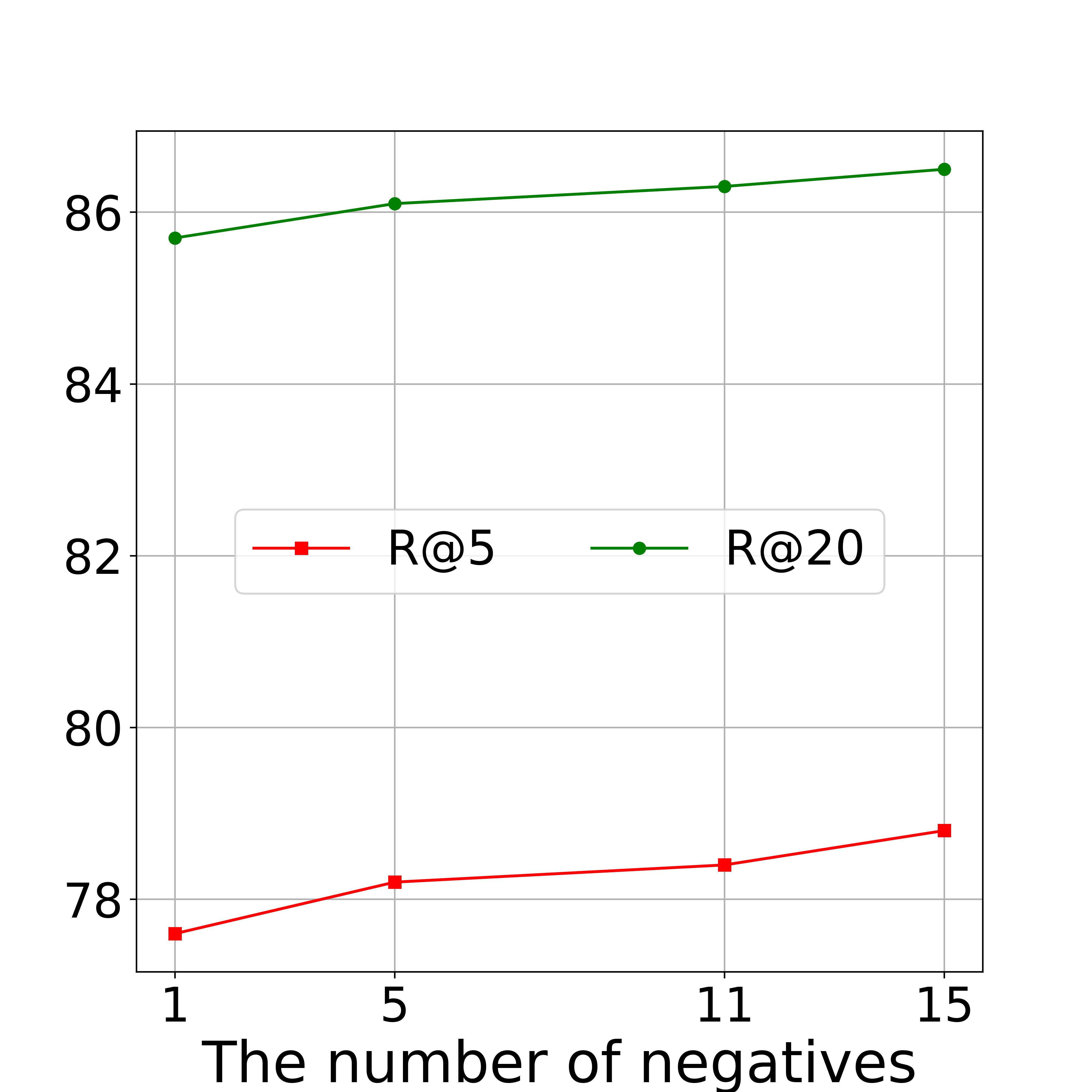}}
    \subfigure[TQA]{
    \includegraphics[width=0.22\textwidth]{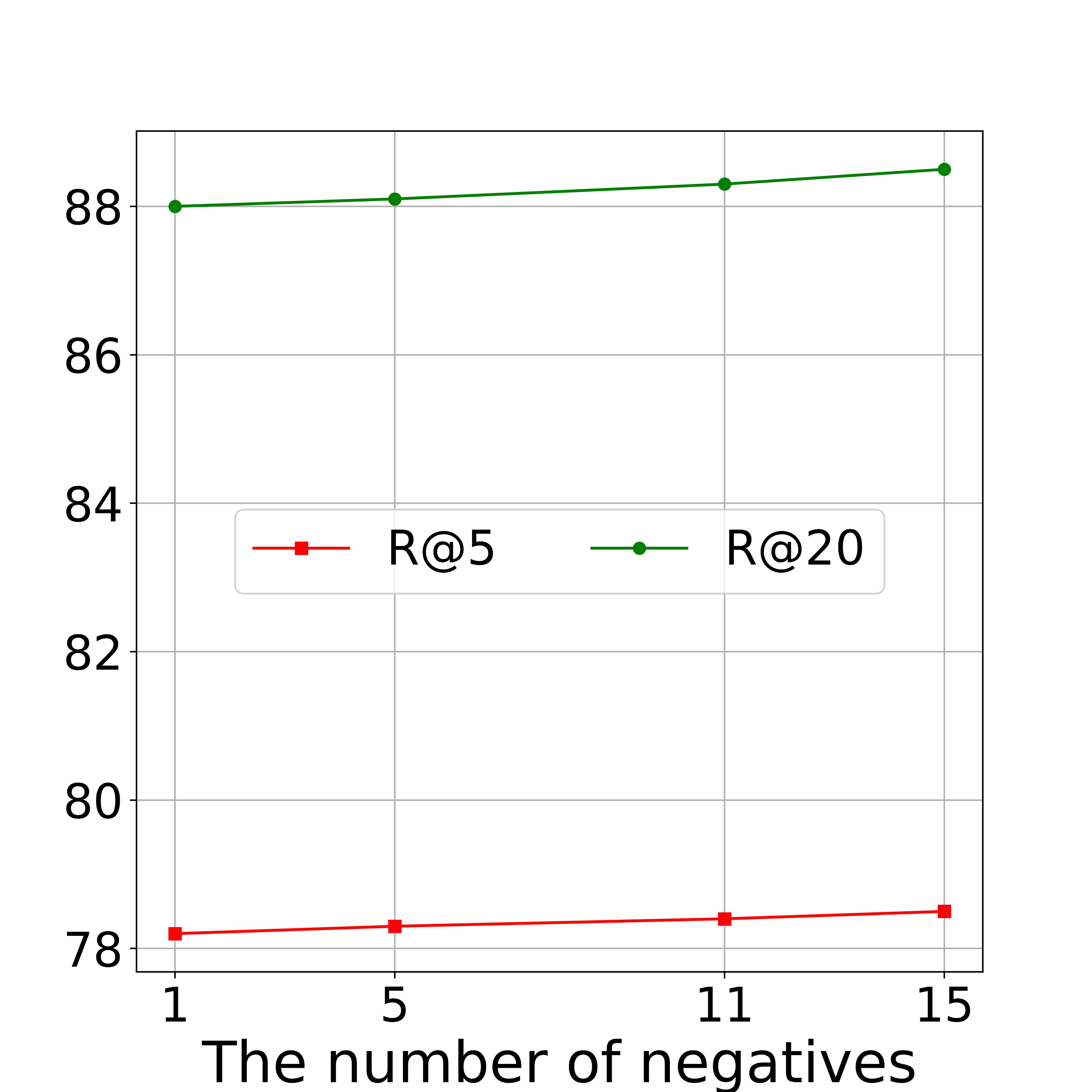}}
    \caption{The impact of negative sample size on the NQ dataset.}
    \label{fig:sampled_size}
    \vspace{-0.6cm}
\end{figure}

\paragraph{Impact of negative sample size.} We further investigate the impact of negative sample size $k$ on retrieval performance using the AR2 model. We vary $k$ in the range of $\{1,5,11,15\}$ and conduct experiments on the NQ and the TQA datasets. As depicted in Figure~\ref{fig:sampled_size}, retrieval performance consistently enhances with the increasing number of negatives, verifying the significance of negative sample size in improving performance. These experimental results align with findings from RocketQA, which also suggest that increasing the number of negatives contributes to better retrieval performance.

\begin{figure}[h]
    \centering
    \vspace{-0.2cm}   
    \includegraphics[width=0.32\textwidth]{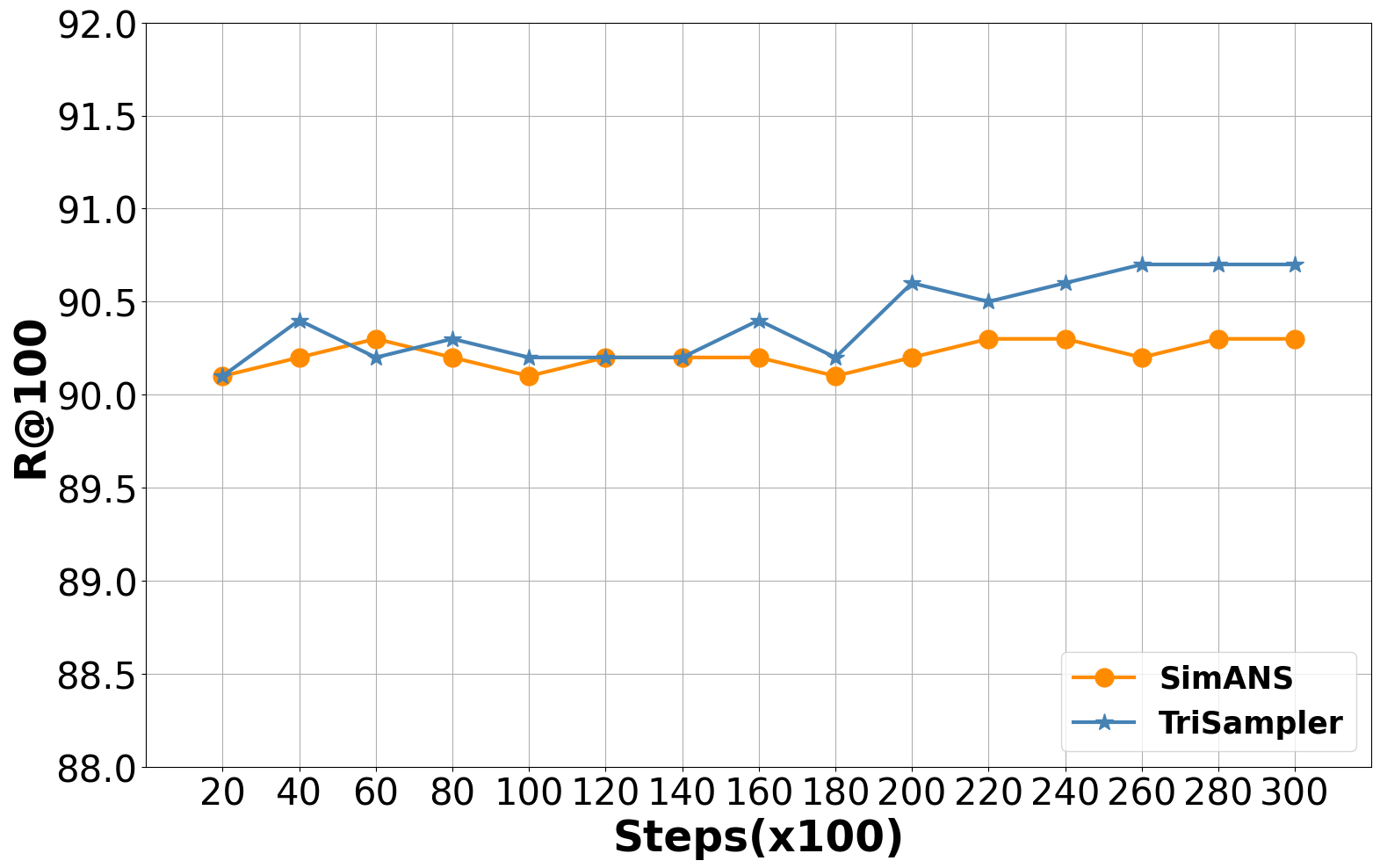}
    \caption{Training convergency curves comparison between SimANS and \smodel on the NQ dataset.}
    \label{fig:convergence}
    \vspace{-0.5cm}
\end{figure}

\paragraph{Training efficiency comparison.} To explore the training efficiency of \model, we test the wall-clock time cost, including the cost of training per batch $\text{Cost}_{D}$ and the cost of training instances construction $\text{Cost}_{C}$. As shown in Table~\ref{tab:efficiency}, it is obviously observed that the training cost of \smodel is slightly higher compared with SimANS. Although \smodel requires more time to construct training instances, the cost is distributed across $t=2000$ training steps, resulting in a per-batch cost of $\text{Cost}_{C/t} = 0.055s$. Thus, the overall cost for training each batch has increased only slightly. However, the total training time to reach optimal performance is reduced because \smodel achieves faster convergency (See Figure~\ref{fig:convergence}). To sum up, \smodel demonstrates improved efficiency gains in comparison to SimANS.

\begin{table}[h]
    \centering
    \small
    \renewcommand{\arraystretch}{1.15}
    \setlength{\tabcolsep}{1.8mm}{
    \begin{tabular}{c|cccc}
        \hline
         Method & Cost$_{D}$ & Cost$_{C}$ & Cost$_{C/t}$ & Cost$_{all}$  \\
        \hline 
        AR2+SimANS &  2.9s & 85s & 0.043s & 2.943s \\
        AR2+\model &  3.0s & 110s & 0.055s  & 3.055s \\   
        \hline
    \end{tabular}}
    \caption{Training efficiency comparison on the NQ dataset.}
    \label{tab:efficiency}
    \vspace{-0.5cm}
\end{table}

\label{sec:experiment}

\section{Conclusion}

In this paper, we investigate the fundamental principle that negative sampling should satisfy in dense retrieval. We first analyze negative sampling from the perspective of objective and subsequently propose a general principle to guide negative sampling, termed \textit{the quasi-triangular principle}. This principle advocates for the confinement of sampled negatives within a region reminiscent of a triangle. Capitalizing on this principle, we propose a simple and effective negative sampling method \smodel to sample more informative negatives within the designated constrained region. Experiments conducted across four benchmark datasets demonstrate the efficacy of \model, showcasing its capacity to deliver superior retrieval performance when compared to alternative strategies.

\label{sec:conclusion}

\section{Acknowledgments}
This work is supported by the Technology and Innovation Major Project of the Ministry of Science and Technology of China under Grant 2022ZD0118600, Natural Science Foundation of China (NSFC) 62276148, Tsinghua University Initiative Scientific Research Program 20233080067, the New Cornerstone Science Foundation through the XPLORER PRIZE.

\bibliography{aaai24}

\end{document}